\documentstyle[10pt,fleqn,twocolumn,epsfig]{article}
\begin{document}
\title{The
       self-organized
       multi-lattice
       Monte Carlo
       simulation}
 \author{     Denis Horv\'ath and Martin Gmitra
              \\
              Department of Theoretical
              Physics and Astrophysics, \\
              University of
              P.J.\v{S}af\'arik,                    \\
              Moyzesova 16, 040 01 Ko\v{s}ice,                    \\
              Slovak Republic                                     \\
       }

 \date{}
 \maketitle

 \begin{abstract}
The self-organized Monte Carlo simulations of 2D Ising ferromagnet on the square lattice are performed.
The essence of devised simulation method is the artificial dynamics consisting of the single-spin-flip 
algorithm of Metropolis supplemented by the random walk in the temperature space. The walk is biased to the 
critical region through the feedback equation utilizing the memory-based filtering recursion instantly 
estimating the energy cumulants. The simulations establish that the peak of the temperature probability 
density function is located nearly the pseudocritical temperature pertaining to canonical equilibrium. 
In order to eliminate the finite-size effects, the self-organized approach is extended to multi-lattice 
systems, where feedback is constructed from the pairs of the instantaneous running fourth-order cumulants
of the magnetization. The replica-based simulations indicate that several properly chosen steady  
statistical distributions of the self-organized Monte Carlo systems resemble characteristics 
of the standard self-organized critical systems. 
 \end{abstract}

 \vspace{0.3cm}


 \vspace{0.2cm}

%----------------------------------------------------------------------
\section{Introduction}
%----------------------------------------------------------------------
The Monte Carlo (MC) simulation 
methods
are nonperturbative 
tools of the statistical 
physics developed hand in hand 
with increasing power 
of nowadays computers.
The benchmark for testing 
of MC algorithms 
represents the exactly 
solvable Ising
spin model  
\cite{ExactSol}. 
Between algorithms applied 
to different 
variants of this model, 
the using of the single-spin-flip  
Metropolis algorithm  
\cite{Binder1988,Metropolis}
prevails due to its simplicity. 
Nevertheless the principal  
problems have emerged 
as a consequence of 
the accuracy and efficiency 
demands especially for the 
critical region.

As a consequence of this
several methods 
enhancing MC efficiency 
have been considered. 
The procedures of great significance are:  
finite-size-scaling relations and renormalization
group based 
algorithms \cite{RGMCs} - \cite{RGMCs3},
cluster algorithms \cite{Swendsen87,Wolff89} 
lowering the critical 
slowing down,
histogram and reweighting techniques 
\cite{Ferrenberg} 
interpolating the stochastic data,
and multicanonical ensemble methods \cite{Berg92}
overcoming the tunneling 
between coexisting phases 
at 1st order transitions. 
Even with the mentioned 
modifications and related 
MC versions, 
a laborious and 
human assisted work is needed until 
a satisfactory 
accuracy 
of results is achieved.
This is trivial 
reason 
why utilization 
of the self-organization 
principles 
have attracted recent attention of MC community.

In the present paper we deal 
with the combining  
of the self-organization 
principles and MC dynamics.
Of course, this effort has 
a computational rather 
than physical impact. 
Our former aim was to design 
the temperature scan seeking for the   
position of the critical points 
of the lattice spin systems. 
But this original 
aim has been later 
affected by the general 
empirical idea 
of the self-organized 
criticality (SOC)~\cite{Bak88}, 
originally proposed 
as a unifying theoretical
framework describing a vast 
class of systems evolving 
spontaneously 
to the critical 
state. 
The SOC examples are
sand pile, 
forest-fire 
\cite{Drossel92} 
and game-of-life 
\cite{Alstrom94} models.
The SOC property 
can be defined 
through the scale-invariance 
of the steady state asymptotics 
reached by the {\em probability 
density functions} (pdf's)
constructed for spatial 
and temporal 
measures of dissipation events called {\em avalanches}.
The dynamics of standard SOC systems  
is governed by the specific 
nonequilibrium critical exponents
linked by the scaling relations \cite{Tang88} 
in analogy to standard phase transitions.

Notice that dynamical rules 
of standard SOC systems 
are functions of the microscopic parameters uncoupled   
to the global control parameters.
On the contrary, 
in the spin systems 
governed 
by the single-spin-flip Metropolis dynamics, 
the spins 
are flipped according to 
prescriptions  
depending on their neighbors, 
but also on the global or external control 
parameters, like temperature    
or selected parameters 
of the Hamiltonian.
The modification, by which the MC 
spin dynamics should be affected 
to mimic the SOC properties, 
is discussed in \cite{Sornette}. 
In agreement with \cite{Kadanoff}, 
any such modification needs the support 
of nonlinear feedback mechanism 
ensuring the critical steady 
stochastic state. 
It is clear that the feedback 
should be defined 
in terms of MC instant estimates 
of the statistical averages  
admitting the definition 
of critical state.

The real feedback model called  
{\em probability-changing 
cluster algorithm} \cite{Tomita2001} 
was appeared without 
any reference to the general 
SOC paradigm.   
The alternative model
was presented in \cite{Fulco99}, 
where temperature of the ferromagnet 
Ising spin system was driven 
according to recursive formula 
corresponding 
to general statistical 
theory \cite{Robbins51}. 
This example was based 
on the mean magnetization 
leading to series 
of the temperature moves 
approaching the magnetic 
transition. 
Despite 
the success 
in the 
optimized localization 
of critical temperature 
of the Ising ferromagnet,
the using of term SOC seems to be 
not adequate for this case. 
The reason is the 
absence of the analysis 
of spatio-temporal aspects 
of MC dynamics, 
which can be 
considered as a 
{\em noncanonical equilibrium} 
\cite{Grandi99,Buonsante2002}, 
due to residual 
autocorrelation 
of the sequential MC sweeps.

Regarding the above 
mentioned approaches, 
the principal 
question arises, 
if the MC supplemented by feedback, 
resembles or really 
pertains to branch 
of the standard SOC models 
which are well-known from 
the Bak's original 
paper \cite{Bak88} 
and related works.

The plan of our paper 
is the following.
Sec.2 is intended to generalization 
of the averaging
relevant for the implementation of 
the self-organization principles.
The details 
of feedback construction 
based on the 
temperature 
gradient of specific heat 
are discussed in Sec.3. 
These proposals are supplemented 
by the simulation 
results carried 
out for 2D Ising ferromagnet.
The details 
of the multi-lattice self-organized 
MC simulations,
stabilizing true critical temperature 
via fourth-order magnetization 
cumulants, are presented in Sec.4.  
The oversimplified mean-field 
model of self-organized MC algorithm 
is discussed in Sec.5. 
Several universal 
aspects of the 
self-organized MC dynamics 
are outlined 
by replica simulations in Sec.6. 
Finally, the conclusions 
are presented.

\section{The running averages}

As we already mentioned 
in introduction, 
the mechanism by which many body system
is attracted to 
the critical
(or pseudocritical) point 
should be mediated by 
the feedback depending
on the instantaneous  
estimates of 
the statistical 
averages. 
In this section
we introduce the 
running averages 
important 
to construct 
the proper feedback rules.

Consider the MC simulation 
generating 
the sequence 
of configurations
$\{\,X_{t'},\, t'=1,2,\ldots,\,t\,\}$
according to importance 
sampling update prescription 
of Metropolis 
\cite{Metropolis} 
producing  
the canonical 
equilibrium Boltzmann 
distribution 
as a function of the constant 
temperature $T$.
The estimate of the canonical 
average
$\langle A\rangle_t$
of some quantity 
$A$
\begin{equation}
\langle A \rangle_{t,T}
= \sum_{t'=1}^{t'=t}
w_{t}
A_{t'}\,\,,
\qquad
w_{t}=
\frac{1}{t}\,.
\label{Eqeta}
\end{equation}
can simply calculated
from 
the series of $t$ sampled
real values $A_{t'} \equiv A(X_{t'})$
reweighted by $w_{t}$
ensuring the trivial 
normalization
\begin{equation}
\sum_{t'=1}^{t}
w_{t} =
w_t \sum_{t'=1}^{t} 1 
 =1\,\,\,.
\label{norm1}
\end{equation}
The summation
given by 
Eq.(\ref{Eqeta}) is equivalent
to recurrence
\begin{equation}
\langle A
\rangle_{t,T} =
(1-w_{t}) \langle A\rangle_{t-1,T} +
w_{t} A_{t}
\label{Eq3}
\end{equation}
showing how the average 
changes 
due
to terminal 
contribution 
$A_{t}$.
Consider 
the 
generalized 
averaging, where
$w_{t}$
is replaced 
by the constant
parameter $0<\eta\ll 1$ 
which is independent 
of
$t$:
\begin{equation}
\langle A
\rangle_{\eta,t,T} =
(1-\eta)
\langle A\rangle_{\eta,t-1,T} +
\eta A_{t}\,.
\label{Eq31}
\end{equation}
The consequence
of setting 
$w_{t}\rightarrow \eta$
is the convolution
\begin{equation}
 \langle
 A \rangle_{\eta,t,T} =
 \sum_{t'=-\infty}^{t}
 w_{\eta,t,t'}
 A_{t'}\,\,.
\label{Aver2}
 \end{equation}
defined 
by the modified weights
\begin{equation}
 w_{\eta,t,t'}=
 \eta\,
(1-\eta)^{t-t'}\,.
\label{gamma1}
\end{equation}
which undergo
to normalization
\begin{equation}
\sum_{t'=-\infty}^{t}
w_{\eta,t,t'} = 1\,.
\end{equation}
For finite initial choice $t'=1$ it yields 
\begin{equation}
\sum_{t'=1}^{t'=t}
w_{\eta,t,t'} = 1 - (1 - \eta)^{t}\,,
\end{equation}
where $(1-\eta)^t \ll 1$
if the time of averaging 
is sufficiently large 
$\eta\ll 1 \ll t$.
It should be remarked that 
the generalized 
averages 
labeled by $\langle A \rangle_{\eta,t,T}$
are equivalent to 
gamma filtered \cite{Principe2000}
fluctuating inputs $A_{t}$. 
Note that 
the term 
{\em gamma} 
originated 
from the analytic form 
of the weight 
$w_{\eta,t,t'}$. 
The filtering represents 
the application of the  
selection principle 
suppressing the information 
older than the memory 
depth $\propto 1/\eta$.

\section{The specific
         heat
         feedback}

To attain the critical 
region self-adaptively we
construct the temperature 
dependent
feedback changing 
the temperature
in a way 
enhancing extremal fluctuations.
The running estimates 
of averages 
are necessary to predict 
(with share of the uncertainty)
the actual system position 
and the course 
in the phase diagram 
leading to critical point.

The pseudocritical 
temperature
$T_{\rm c}(N)$
of some 
finite system
consisting of $N$ 
degrees of 
freedom
is defined 
by the maximum
$C(N,T_{\rm c}(N))$
of the specific heat $C(N,T)$.
To form an attractor
nearly $T_{\rm c}(N)$,
we propose the following  
dynamics of the 
temperature random walker
\begin{equation}
T_{t+1}= T_{t}
+  r_t
\,
\Delta\,\,
\mbox{sign}(
F^{C}_t)
\label{EqFc}
\end{equation}
biased by the gradient
\begin{equation}
F^{C}_t=
\left.
\frac{\partial C}{\partial T}\,\right|_{T=T_{t}}\,,
\label{gradC1}
\end{equation}
where $C$ 
is MC estimate of the specific heat.
From that 
follows that 
the energy fluctuations 
are controlled 
by the temperature representing 
the additional 
slowly varying degree of freedom.
It is assumed 
here and in further that
$T_t$ remains  constant 
during $N$ random 
microscopic moves
($1$MC step per $N$).
The sign function 
occurring in Eq.(\ref{EqFc})
is used to suppress the extremal
fluctuations of 
$F^{C}_t$ causing the unstable 
boundless behavior of $T_t$.
From several preliminary 
simulations it can be concluded 
that the replacement of sign function 
by some smooth differentiable function  
(e.g. tangent hyperbolic or arcus tangent) 
seems to be irrelevant for keeping 
of smaller dispersion 
of $T_t$.
The non-constant 
temperature steps
are constrained by 
$|T_{t+1}-T_t|<\Delta$
due to 
action 
of the   
pseudorandom numbers $r_t$
drawn from the uniform 
distribution
within the 
interval $(0,1)$. 

Very important  
for further purposes is 
the quasi-equilibrium 
approximation 
$\langle  A  \rangle_t
\simeq
\langle  
A  
\rangle_{\eta,t,T_t}$ 
justified under the restrictions
$t_{{\rm eq},A}\,\eta \ll 1$, 
$t_{{\rm eq},A}\,\Delta \ll 1$, 
where 
$t_{{\rm eq},A}$
is the equilibration time of $A$.  
With help of this 
approximation, 
the running 
averages from Eq.(\ref{Eq31})
can be 
generalized using 
\begin{equation}
\langle A \rangle_{\eta,t,T_t} = 
(1-\eta) 
\langle A \rangle_{\eta,t-1,T_{t-1}} + w_t A_{t,T_t}\,,
\label{Eq31a}
\end{equation} 
which allow the averaging 
under the slowly 
varying temperature.
Here $T_t$ is the
temperature for 
which the last sample
$A_{t,T_t}=A(X_t)$ 
is calculated.

For later purposes 
we also introduce 
the zero 
{\em passage time}
$\tau_{A}$ 
pertaining to $A$.
The time is defined 
as a measure 
of the stage during 
which the
$\mbox{sign}(A_t)$ 
is invariant.
More 
formal definition 
of $\tau_A$
requires the introducing 
of two auxiliary times $t'$, $t''$. 
The first time $t'$ 
defines the instant, where  
\begin{equation}
A_{t'-1}\,A_{t'}\leq 0\,,
\label{signFtC1}
\end{equation}
whereas $t''$ counts the events for which  
\begin{equation}
A_{t'+t''-1}
\, A_{t'+t''}>0\,,
\,\,\,\, 
t'' = 1,2,\ldots \tau_A>1\,.
\label{signFtC2}
\end{equation}
The counting is 
finished for $t''=\tau_A$ if 
\begin{equation}
A_{t'+\tau_A}\,
A_{t'+\tau_A+1}\leq 0\,.
\label{signFtC3}
\end{equation}
To be thorough, 
the conditions
$A_{t'-1}\, A_{t'} \leq 0$,
$A_{t'}\, A_{t'+1} \leq 0$ 
should be assumed 
for definition of $\tau_A=1$.  
Thus, using 
Eqs.(\ref{signFtC1})-(\ref{signFtC3}) 
the original sequence 
$\{A_t\}_{t=1}$ 
can be 
transformed 
to the sequence of passage times 
$\{\tau_A^{(k)}\}_{k=1}$.

From 
the above definition 
it follows
that random walk in temperature 
is unidirectional 
(the sign of $T_{t+1}-T_t$ is ensured)
for $t''=1,2,\ldots, \tau_{F^C}$. 
The arithmetic
average 
of 
$\tau_{F^{C}}$
is related 
to temperature 
dispersion
$\langle  (\delta T)^2 \rangle=
(\Delta^2/3)
\langle
\tau_{{F^{C}}}
\rangle$, 
where 
$
\langle
\tau_{{F^{C}}}
\rangle$
is calculated 
from 
$\{ \tau_{F^C}^{(k)} \}_{k=1}$.  
The standard 
formula
providing $C(T)$  
is the 
fluctuation-dissipation
theorem 
(in $k_{\rm B}$ units)
\begin{equation}
C =
\frac{
\langle E^2\rangle_{t,T}-
\langle E\rangle_{t,T}^2}{T^2 N}\,.
\end{equation}
Here the 
specific heat
is expressed
in terms of the  
energy 
$E$ 
cumulants
$\langle E \rangle_t$,
$\langle E^2 \rangle_t$.
In the
frame 
of the 
quasi-static 
approximation it
is assumed:
$T\simeq T_t$,
$\langle E \rangle_{t,T} 
\simeq
 \langle E \rangle_{\eta, t,T_t}$,
$\langle E^2 \rangle_{t,T}
\simeq \langle
E^2 \rangle_{\eta, t,T_t}$.
Subsequently, 
using
properties of 
energy cumulants 
with 
equilibrium
Boltzmann weights,
the temperature
derivative 
of
$C$ can be 
approximated by
\begin{eqnarray}
&& F^{C}_t 
=
\frac{
\langle E^3 \rangle_{\eta,t,T_t} -
3 
\langle E   \rangle_{\eta,t,T_t}
\langle E^2 \rangle_{\eta,t,T_t}
}{ T_t^4 N }
\nonumber
\\
&&+
2  \frac{\langle E \rangle_{\eta,t,T_t}^3 }{T_t^4 N}
- 2 \frac{ 
\langle E^2 \rangle_{\eta,t,T_t} 
- \langle  E \rangle_{\eta,t,T_t}^2}{T_t^2 N}
\,.
\label{DCDT1}
\end{eqnarray}
%--------- FIGURE ---------
\begin{figure}[!h]
\epsfig{file=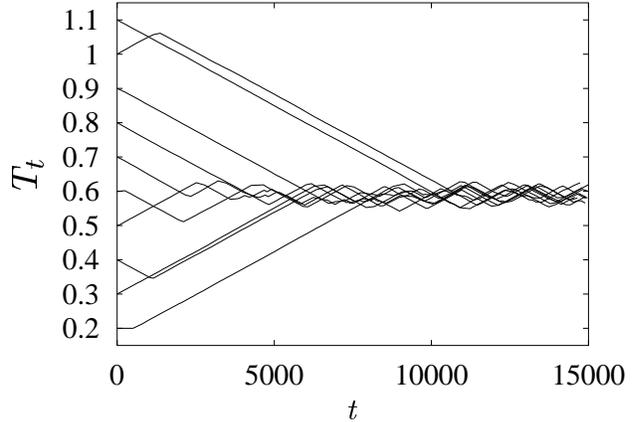,width=8.5cm,angle=0.0}
\caption{
   The
   transient 
   regime 
   of  
   $T_t$
   obtained for several
   different
   initial
   conditions.
   Simulated 
   for
   $L=10$,
   $\eta=  10^{-3}$,
   $\Delta=10^{-4}$
   and identical 
   initial values 
   of cumulants. 
  }
\label{Fig1}
\end{figure}
%--------------------------
%--------- FIGURE ---------
\begin{figure}[!h]
\epsfig{file=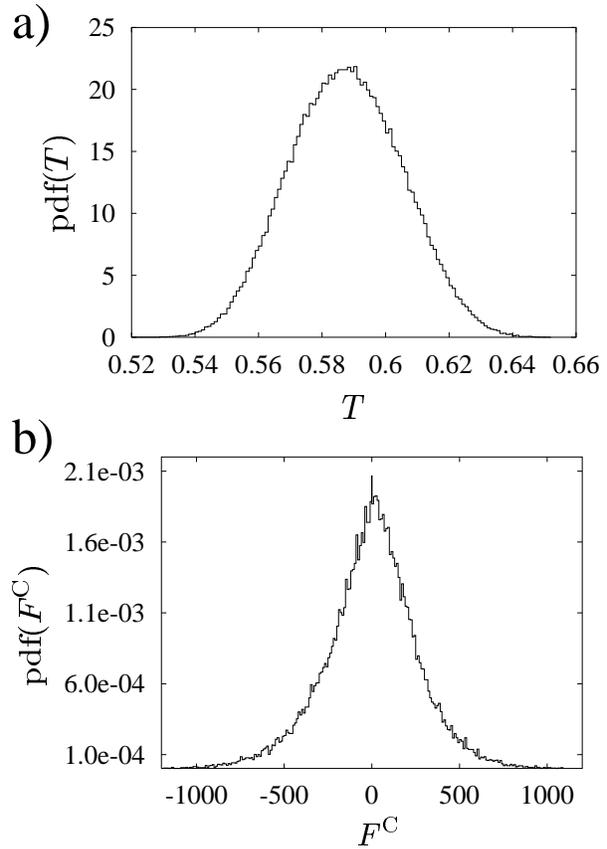,width=8.5cm,angle=0.0}
\caption{
The pdf 
distributions
obtained for 
$\eta=10^{-3}$,
$\Delta=10^{-4}$,
$L=10$:
a)~the pdf of temperature
    with the peak
    located
    nearly
    $T_{\rm c}(N)
    \simeq 0.5868$
    (dispersion
    $0.018$);
b)~the simulation 
   reveals the flat 
   tails of pdf of $F^{C}_t$.
   }
\label{Fig2}
\end{figure}
%---------------------------

The Ising 
ferromagnet 
is simulated in further 
to study the effect of feedback
defined 
by Eqs.(\ref{EqFc}),
(\ref{gradC1})
and (\ref{DCDT1}). 
However, it is
worthwhile to note
that many of the presented 
results are of general relevance.
Given 
the spin system 
$X=\{ s_i\}_{i=1}^N$,  
$s_i=\pm 1/2$ 
placed 
at $N=L^2$ sites
of the square  
$L\times L$ 
lattice with 
the 
periodic boundary
conditions, 
the Ising  
Hamiltonian can be defined 
in exchange coupling units
\begin{equation}
E= - \sum_{\rm nn} s_i s_j\,, 
\end{equation}
where nn means that summation 
running  
over the spin nearest neighbors.

\begin{figure}[!h]
\epsfig{file=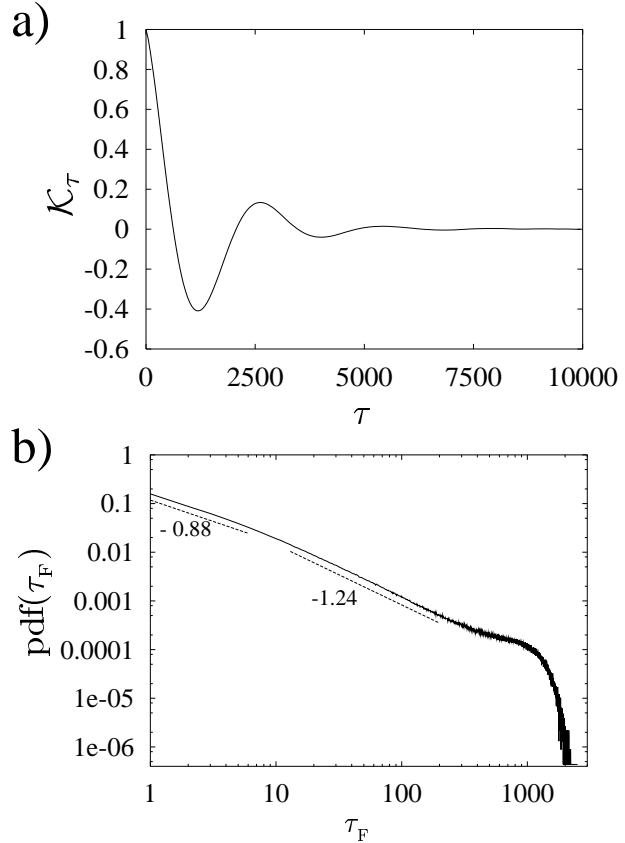,width=8.5cm,angle=0.0}
\caption{
   The
   simulation for
   $L=10$,
   $\eta= 10^{-3}$,
   $\Delta=10^{-4}$;
a)~the
   autocorrelation
   function
   $K_{\tau}$;
b)~the
   log-log
   plot
   of the
   normalized pdf
   of zero
   passage time of
   $\tau_{{F^{C}}}$
   supplemented
   by
   the local 
   slope
   information.
   }
\label{Fig3}
\end{figure}

In general, the dynamics of SOC systems
exhibits two distinct regimes. 
During the {\em transient} regime the proximity of critical 
or pseudocritical point is reached.
The second steady regime 
is called here the 
noncanonical equilibrium in analogy with \cite{Buonsante2002}. 
In this  
regime the attraction to critical point 
is affected 
by the critical noise.
This general 
classification is confirmed 
by our results shown in 
Figs.\ref{Fig1}-\ref{Fig3}.
In Fig.\ref{Fig1} 
we see the stochastic 
paths of $T_t$ pertaining
to different 
initial values 
of energy cumulants
and spin configurations.  
The paths are 
attracted by 
$T_{\rm c}(N)$ 
with some 
uncertainty in noncanonical 
equilibrium.
For the sufficiently narrow 
steady pdf's, 
$T_{\rm c}(N)$ 
can be approximated
by 
$T_{\rm c}(N)
\simeq N_{\rm av}^{-1}
\sum_{t=1}^{N_{\rm av}} T_{t}$,
where $N_{\rm av}$ 
is the number of inputs.
The stationary pdf 
of $T_t$ walk 
is shown 
in Fig.\ref{Fig2}a and  
pdf of $F^{C}_t$ 
with non-Gaussian 
flat tails 
is depicted in Fig.\ref{Fig2}b.

The alternative quantity 
capable for the 
characterization of the
noncanonical 
equilibrium 
is the autocorrelation 
\begin{equation}
{\cal K}_{\tau}=
\frac{1}{N_{\rm av}}
\sum_{t=1}^{N_{\rm av}}
F^{C}_{t}
F^{C}_{t+\tau}\,\,.
\end{equation}
The simulations 
results depicted 
in Fig.\ref{Fig3}a
evidenced that minimum time 
for which the anticorrelation
($K_{\tau}<0$)
occur is of the order  
${\cal O}(\langle\tau_{{F^{C}}}\rangle)$. 
As we see 
from Fig.\ref{Fig3}b,
the power-law 
dependence
can be identified 
within the region 
of vanishing $\tau_{{F^{C}}}$.
More profound 
discussion 
of this fact 
is presented 
in Sec.6.

%--------- FIGURE ---------
%-------------------------------------------------------------------

\section{Multi-lattice simulations} 

In this
section
we try 
to avoid the problem
of finite-size-scaling related 
to true
equilibrium critical
temperature
$T_{\rm c}$ 
and 
critical exponents.
The problem 
is solved
by the
multi-lattice
self-organized
simulations
based on the dynamical rules 
treating the information 
from running averages 
of magnetization. 
The considerations 
are addressed to 
models,
where 2nd 
order phase 
transitions
take place. 
The proposal 
is again 
applied to
2D~Ising 
ferromagnet 
on the square 
lattice.

The quantity
indicating 
deviations
of magnetization  
order parameter
$m_{L}=
(1/L^2)\sum_{\langle ij \rangle} s_i$
from the gaussianity
is the fourth-order 
cumulant
\begin{equation}
U_{L,T}= 1-
\frac{\langle  m_L^4  \rangle_t }{ 3
\langle m_{L}^2
\rangle_t^2 }\,.
\label{UL12}
\end{equation}

The standard way
leading to
true 
$T_{\rm c}=\lim_{N\rightarrow \infty} 
T_{\rm c}(N)$
is the construction 
of the temperature
dependences
$U_{L_l,T}$, 
$U_{L_s,T}$
for 
two lattices
$L_s \neq  L_l$.
Then 
$T_{\rm c}$
follows 
from the condition 
of the 
scale-invariance
\begin{equation}
U_{L_l,T_{\rm c}}=
U_{L_s,T_{\rm c}}\,\,.
\end{equation}
According it 
the self-organized 
multi-lattice 
MC simulation method 
consists of the 
following three 
main points  
repeated in the canonical 
order 
for the  
counter 
$t=1,2,\ldots \,$:

\begin{enumerate}

\item        The 
             performing of 
             $L_l^2$
	     spin flips
             on the lattice 
             indexed by~$l$
             and
             $L_s^2$
	     spin flips
             on 
             the second lattice.
             The flips are 
	     generated for 
             fixed temperature 
             $T_t$. 
             After it, the instant 
             magnetizations (per site)
             $m_{L_l,T_t}$ and 
             $m_{L_s,T_t}$ 
             are calculated.
\item
The update of the cumulants
$\langle m_{L_l}^2 \rangle_{\eta,t,T_t}$,
$\langle m_{L_s}^2 \rangle_{\eta,t,T_t}$,
$\langle m_{L_l}^4 \rangle_{\eta,t,T_t}$,
$\langle m_{L_s}^4 \rangle_{\eta,t,T_t}$
according to Eq.(\ref{Eq31a}),
which yield to the modified 
of definition 
from 
Eq.(\ref{UL12})
\begin{equation}
U_{L,T_t}= 1-
\frac{
\langle  m_L^4  \rangle_{\eta,t,T_t} }{ 3
\langle m_{L}^2
\rangle_{\eta,t,T_t}^2 }\,\,\,.
\label{UL12x}
\end{equation}

\item
The 
temperature shift
\begin{equation}
T_{t+1} =
T_t +
r_t \,\Delta\,
\mbox{sign}(F_{t,ls}^{ U})
\label{Ttt2}
\end{equation}
biased to 
eliminate difference
\begin{eqnarray}
F_{t,ls}^{U}=
U_{L_l,T_t} - U_{L_s,T_t}\,.
\label{CoupliU12}
\end{eqnarray}
If $L_l>L_s$ 
the ordering
of cumulants in $F^U_{t,ls}$
is chosen subject to assumption
\begin{eqnarray}
U_{L_l,T_t} &>&
U_{L_s,T_t}
\quad
\mbox{for}
\,\,\,\,\,\,
T_t<T_{\rm c}\,\,,
\\
U_{L_l,T_t}
&<&
U_{L_s,T_t}
\quad
\mbox{for}
\,\,\,\,\,\,
T_t>T_{\rm c}\,\,.
\nonumber
\end{eqnarray}
\end{enumerate}
Any modification of the Eqs.(\ref{Ttt2}) and (\ref{CoupliU12})
is possible when the preliminary recognition 
of the critical point neighborhood is performed. 
The 
parametric tuning 
recovers that 
stabilization 
of the noncanonical equilibrium
via feedback
$F_{t,ls}^{U}$ 
requires 
smaller
$\eta$ and 
$\Delta$
than single-lattice 
simulations based 
on action of $F_{t}^{C}$.
The Eq.(\ref{Ttt2})
can be generalized  for
$n_{\rm r}$
lattices, i.e. for
$n_{\rm rp}=
\frac{n_{\rm r}}{2} ( n_{\rm r}-1 )$
competing lattice  
pairs labeled 
by $l,s$:
\begin{eqnarray}
T_{t+1}
= T_t+
\frac{r_t\, \Delta}{n_{\rm rp}}\,
\sum_{l<s=1}^{n_{\rm r}\times n_{\rm r}}
\mbox{sign}
\left(\,
F_{t,ls}^{U}
\,\right)\,,
\label{three111}
\end{eqnarray}
where $1/n_{\rm rp}$ 
term 
rescales additive
contributions.

\begin{figure}[!h]
\epsfig{file=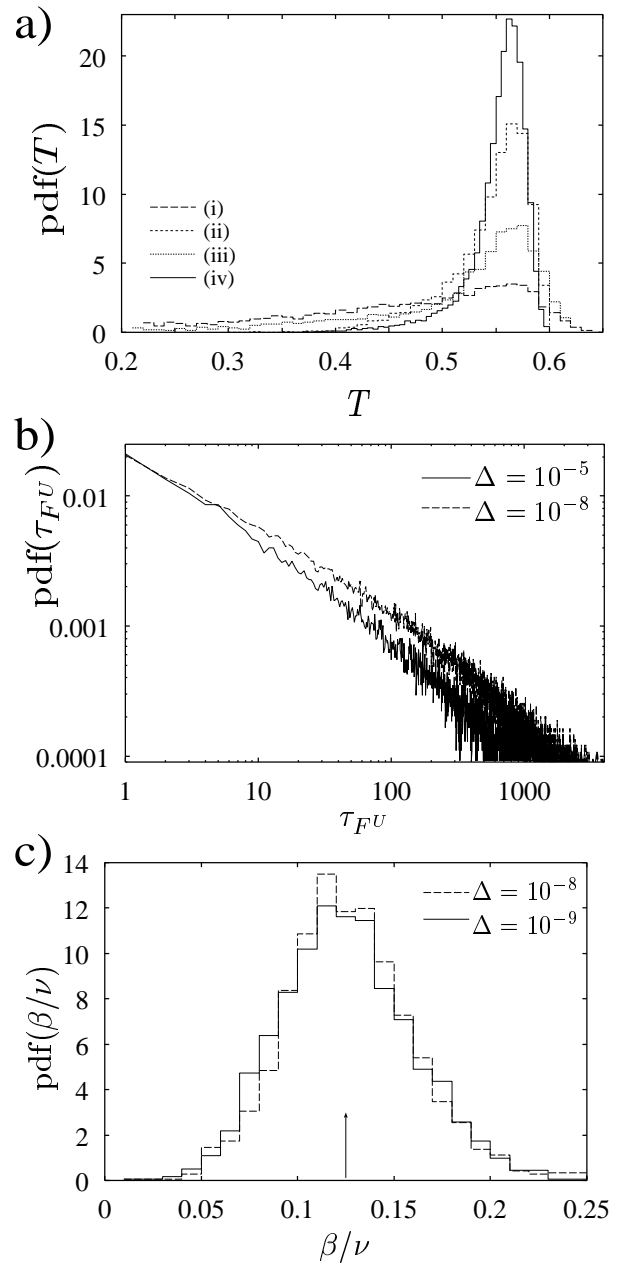,width=8.0cm,angle=0.0}
\caption{
The stationary
statistics of lattices coupled by
the fourth-order 
magnetization
cumulants.
a)~The pdf's
of temperature
obtained for
parameters
$\eta=10^{-3}$,
$\Delta=10^{-4}$.
$n_{\rm r}=2$
for 
$L_1=10$, $L_2=20$~[see~(i)].
In that case additional 
low 
temperature
bound
$T_t>0.1$
is used to confine dynamics into the region $T_t>0$.
For 
$n_{\rm r}=3$,
$L_1=10$, 
$L_2=16$,
$L_3=20$~(ii)
the stabilization bound
is not necessary.
For  $\eta=10^{-4}$
and for 
$\Delta=10^{-5}$
the location of $T_{\rm c}$
is much
better~[
see $n_{\rm r}=2$~(iii)
and $n_{\rm r}=3$~(iv)];
b)~The 
   comparison
   of pdf's
   of 
   the
   zero  
   passage 
   time  
   $\tau_{{F^{U}}}$
   of 
   two coupled 
   lattice
   systems 
   of sizes $L_1$ and $L_2$. 
    Calculated 
    for
   $\eta=10^{-4}$,
   $\Delta=10^{-5}$
   and
   $\eta=10^{-4}$,
   $\Delta=
   10^{-8}$.
   The log-log 
   plot of 
   pdf's
   results in  
   the slope
   $-0.671$
   in the region 
   of vanishing 
   $\tau_{{F^{U}}}$. 
   For the middle 
   $\tau_{{F^{U}}}$ 
   region 
   the  
   slope 
   is $-0.59$; 
c)~pdf's
   of the 
   effective
   critical
   index
   $(\beta/\nu)_{\eta,t}$
   obtained
   for  
   parametric
   choices
   from~b).
   The 
   arrow 
   indicates  
   the 
   of exact 
   $T_{\rm c}$.}
\label{Fig4}
\end{figure}

The presented  
method also 
offers the 
continuous 
checking 
of estimated critical exponents. 
It comes from the 
standard assumption 
that 
canonical equilibrium magnetization 
exhibits 
critical scaling
$\langle |m_{L_j}|
\rangle_t
=
L_j^{-\beta/\nu}
f_{\rm m}
\left( L_j
(T_t-T_{\rm c})\right)$, 
where $f_{\rm m}(\cdot)$
is the scaling 
function and 
$\beta/\nu$
is the ratio of 
magnetization
($\beta$) 
and the correlation length
($\nu$) critical exponents,
respectively.
If the temperature 
fluctuates
nearly $T_{\rm c}$, 
the equilibrium 
finite-size-scaling
relation 
changes to
$\langle |m_{L_j}|
\rangle_{\eta,t,T_t=T_{\rm c}}
\simeq
L_j^{-\beta/\nu}\,
f_{\rm m}(0)$.
For 
$n_{\rm r}\geq 2$
lattices 
and 
sufficiently small 
$\eta$, $\Delta$,
the following 
arithmetic average can be 
defined
\begin{equation}
\left(
\frac{\beta}{\nu}
\right)_{\eta,t}
=
\frac{1}{n_{\rm rp}}\,
\sum_{l<s=1}^{ 
n_{\rm r} \times n_{\rm r}}
\frac{\ln \left(
\frac{
\langle  | m_{L_l} |
\rangle_{\eta,t,T_t}}{
\langle  | m_{L_s} |  
\rangle_{\eta,t,T_t}}
\right)}{
\ln \left(
\frac{L_s}{L_l}
\right)}\,.
\label{Eqindex1}
\end{equation}
Similar to treatment 
of the steady temperature fluctuations, 
the quantity $\beta/\nu
\simeq
N_{\rm av}^{-1}
\sum_{t=1}^{N_{\rm av}}
(\beta/\nu)_{\eta,t}$ can be defined. 
The simulations
carried out for cases  
$n_{\rm r}=2,\,3$
are compared
in Fig.\ref{Fig4}.
In agreement with 
expectation,
the localization
of $T_{\rm c}$
for
$n_{\rm r}=3$ with recursion taken from 
Eq.(\ref{three111})
is much subtle 
than for
$n_{\rm r}=2$.
In addition, 
the statistics of
$(\beta/\nu)_{\eta,t}$
is weakly
depending on 
$\eta$ and $\Delta$,
which seems to be
logical due to universality of 
exponents in the 
canonical equilibrium limit 
($T_{\rm c}={\rm const}$).

%--------- FIGURE ---------
%--------------------------------

The 
$n_{\rm r}=2$ 
simulations
applied for 
$L_1=10$, 
$L_2=20$,
$\eta=10^{-4}$,  
$\Delta =10^{-8}$
leads 
to the 
noncanonical equilibrium, 
for which the 
temperature average 
is associated 
with estimate 
$T_{\rm c}\simeq 
0.5667$
of the
exact value
$T_{\rm c}^{\rm ex}
\simeq
[2\ln(1+\sqrt{2}]^{-1}
\simeq 0.56729$.
The ratio 
$\beta/\nu 
\simeq 0.122$ 
approximates the 
exact index  
$(\beta/\nu)^{\rm ex}
=0.125$. Much 
slower walk for 
$\Delta =10^{-9}$ 
provides 
only a insufficient 
improvement 
of the previous 
results. 
More 
appealing are 
estimates
$T_{\rm c}=0.5673$, 
$\beta/\nu
\simeq 0.123$
obtained for 
$n_{\rm r}=3$,
$L_1=10$, 
$L_2=20$, 
$L_3=30$,
$\eta=10^{-4}$, 
$\Delta=10^{-9}$, 
$N_{\rm av}=5\times 10^8$ 
with balance of cumulants attained for  
$U_{L=10,20,30,T_{\rm c}} \simeq 0.61$.
Note that $\beta/\nu$ does not 
change substantially 
[$\beta/\nu\simeq 0.123(5)$] 
if estimated from 
averages
$\langle |m_{L_1=10}|\rangle_{\eta,t,T_t}=0.37$,
$\langle |m_{L_2=20}|\rangle_{\eta,t,T_t}=0.34$,
$<|m_{L_3=30}|\rangle_{\eta,t,T_t}=0.33$.

\section{The
         mean-field
         analysis of
         algorithm}

\begin{figure}[!h]
\epsfig{file=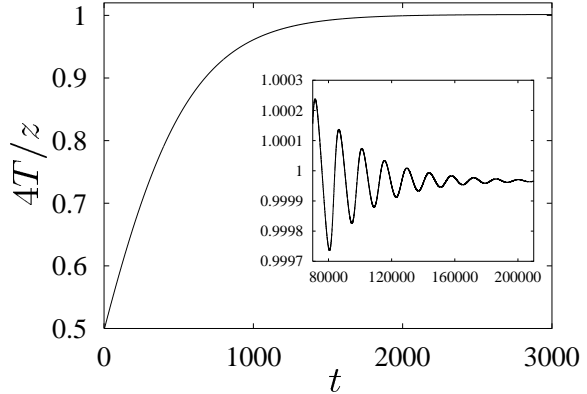,width=8.0cm,angle=0.0}
\caption{
    The 
    numerical 
    solution of 
    differential equations 
    Eqs.(\ref{Expm})
    and (\ref{mmc1}) 
    is presented.
    The transient
    dynamics
    of the  
    temperature
    obtained for 
    $m_{\rm c}
    =0.005$,
    $\alpha=10^{-3}$
    and
    initial
    conditions
    $T|_{t=0}
     = 0.125$,
    $m|_{t=0}=0.05$\,.
    }
\label{Fig5}
\end{figure}

In this section 
we present 
calculations 
aimed to understand  
how the attractivity 
of critical point arises and 
how the noncanonical equilibrium 
is attained by means of feedback. 
Only a rough approximation of the 
complex simulation process 
is considered, where
spin degrees of freedom 
are replaced by the 
unique magnetization~(per site) 
term $m(t)$. 
Furthermore, 
it assumes that 
the selected central spin $s(t)$ 
flips in a mean field 
created by its 
$z$ neighbors. 
Let $\pi(t)$ 
denotes the probability 
of the occurrence of 
$s(t)=1/2$ state, 
then the probability of 
$s(t)=-1/2$ is 
$1-\pi(t)$.
The master equation for 
$\pi(t)$ can be written in the form
\begin{equation}
\frac{{\rm d}\pi}{{\rm d}t} =
(1-\pi)
W^{\mbox{\tiny $[-+]$}} -
\pi
W^{\mbox{\tiny $[+-]$}}\,.
\label{Eqdpi1}
\end{equation}
The 
Glauber's \cite{Glauber63} 
heat bath dynamics with 
the transition probabilities  
$W^{\mbox{\tiny $[-+]$}}$ 
and 
$W^{\mbox{\tiny $[+-]$}}$ 
between states 
$s(t)=\pm 1/2$ 
is preferred 
in comparison 
to non-differentiable Metropolis 
form due to analyticity arguments 
relevant for formulation by means of differential equation. 
Within  the mean-field 
approximation it can 
be assumed 
\begin{eqnarray}
W^{\mbox{\tiny $[+-]$}}(t)
&=&
\frac{1}{2\tau_{\rm f}}
\,
\left[
1 -
\tanh
\left(
\frac{z m(t)}{2 T(t)}
\right)
\right]
\,,
\label{WWW12}
\\
W^{\mbox{\tiny $[-+]$}}(t)
&=&
\frac{1}{2\tau_{\rm f}}
\,
\left[
1 +
\tanh
\left(
\frac{z m(t)}{2 T(t)}
\right)
\right]
\,.
\nonumber
\end{eqnarray}
In the above expression
$\tau_{\rm f}$ is the time 
associated with the spin flip process. 
The  expression 
takes into account $\pm z m(t)$ variations 
of energy 
belonging to flips 
from $s=\pm 1/2$ 
to $s=\mp 1/2$
given by 
the effective 
single-site 
Hamiltonian 
$-z\, s(t)\,m(t)$.  
Assuming that 
$\pi=m+1/2$ and using Eqs.(\ref{Eqdpi1}), 
(\ref{WWW12}) we obtain
\begin{eqnarray}
&&
\frac{{\rm d}m}{{\rm d} t}
=
\frac{1}{\tau_{\rm f}}
\left[
\frac{1}{2} 
\tanh
\left(
\frac{z m}{2 T}
\right)
- 
m
\right]
\label{Expm}
\\
&&
\simeq \frac{1}{\tau_{\rm f}}
\left[
\frac{z - 4 T}{ 4 T } m -
\frac{z^3 m^3}{ 48 T^3} +
{\cal O}(m^5)
\,
\right]
\,.
\nonumber
\end{eqnarray}
Subsequently, the feedback differential equation 
of the temperature variable
is suggested in the form  
\begin{equation}
\frac{{\rm d}T}{{\rm d}t} =
\alpha \left(
\, m^2 - m_{\rm c}^2
\,\right) \equiv F^m(t)\,,
\label{mmc1}
\end{equation}
where $m_{\rm c}$ is the 
"nucleation" parameter 
of the ferromagnetic
phase, and $\alpha>0$
is the constant parameter.
Unlike the works \cite{Fulco99,Robbins51}, 
where feedback consisting of $|m|$ 
term is considered, 
the $m^2$ is absorbed 
to the feedback $F^m(t)$ 
proposed here 
to ensure 
the analyticity.
For $m^2>m_{\rm c}^2$ 
the temperature
increases, 
whereas 
$m^2<m_{\rm c}^2$ 
leads to the cooling.
The  stationary 
solution of Eqs.(\ref{Expm})
and (\ref{mmc1}) is
\begin{equation}
m_{\rm c} = \frac{1}{2} \tanh
\left(
\frac{z  m_{\rm c}}{2 T} \right)\,.
\label{Eqmc1}
\end{equation}
In the limit of vanishing $m_{\rm c}$,  
the solution of
Eq.(\ref{Eqmc1})
can be written 
in terms 
of the inverse
Taylor series 
in $m_{\rm c}$
\begin{equation}
T=
\frac{z}{4}\left(1-\frac{4 m_{\rm c}^2}{3} \right) +
{\cal O}(m_{\rm c}^4)\simeq T_{\rm c}^{\rm MF}\,,
\label{nearcri1}
\end{equation}
where $m_{\rm c}=0$   
corresponds 
to the known 
mean-field
critical temperature
$T_{\rm c}^{\rm MF}=z/4$.
Small negative shift of stationary  
$T$ from Eq.(\ref{nearcri1})
caused by 
$m_{\rm c}\simeq 0$  
corresponds 
to Fig.\ref{Fig5} 
including the numerical 
solution of Eqs.(\ref{Expm}) and 
(\ref{mmc1}). 

%--------- FIGURE ---------
%-----------

\section{The
         comparison 
	 of
         MC
         and
         SOC
         dyna\-mics.}

\begin{figure}[!h]
\epsfig{file=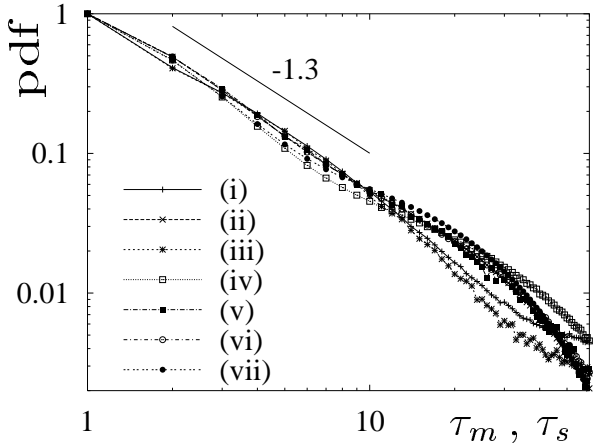,width=8.0cm,angle=0.0}
\caption{
The comparison 
of  
stationary 
pdf's of zero 
passage times 
$\tau_{s}$ 
and $\tau_{m}$ 
of 
different 
lattice sizes 
and 
different 
feedbacks.
Simulated for 
$\eta=10^{-4}$ and $\Delta=10^{-5}$. 
The figure 
shows 
rescaled pdf's 
of different 
systems 
subject to settings 
ordered to 
three-component tuples:  
[feedback type, 
(quantity leading 
to $\{\tau_A^{(k)}\}$ 
sequence), system size]:
(i),~$F^C_t$, ($A=m_L$), $L=10$;
(ii),~$F^U_t$, ($A=s_i$), $L=10$; 
(iii),~$F^U_t$, ($A=m_L$), $L=10$;
(iv),~$F^U_t$, ($A=s_i$), $L=10$;
(v),~$F^C_t$, ($A=s_i$), $L=200$;
(vi),~$F^C_t$, ($A=s_i$), $L=50$; 
(vii),~$F^C_t$, ($A=s_i$), $L=4$;}
\label{Fig6}
\end{figure}

\begin{figure}[!h]
\epsfig{file=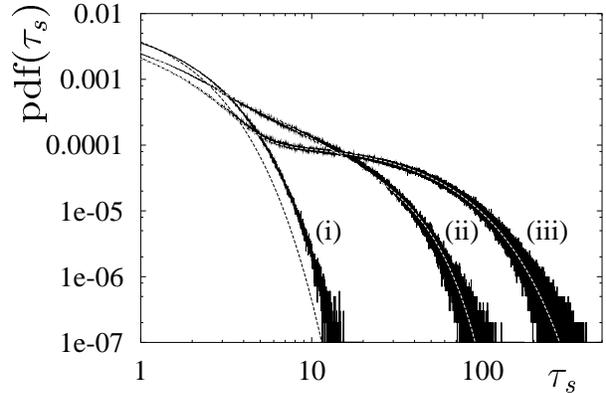,width=8.0cm,angle=0.0}
\caption{
The pdf
of the spin passage
times
obtained
for 
$L=10$,
$\eta=10^{-4}$,
$\Delta= 10^{-6}$.
The MC simulations for  
fixed 
temperatures
(i)~$T=4\,
T_{\rm c}(N)$~({\it para-phase});
black-dashed
line corresponding 
to
${\cal P}_{\tau_{s}}$ 
from Eq.(\ref{Ptpara1});
(iii)~$T=0.8 \,
T_{\rm c}(N)$~({\it ferro-phase}).
The white-dashed line
is the
fit of the "bimodal"
pdf
${\cal P}_{\tau_{s}}$.
Compared with self-organized
MC simulations~(ii),
where 
feedback 
$F_t^{C}$
yields 
to the 
power-law 
behavior 
with exponent from 
Eq.(\ref{expphi1}).
}
\label{Fig7}
\end{figure}

In the section we discuss the universal aspects of 
the non-equilibrium 
self-organized MC dynamics 
applied to the canonical Ising model.
As is already mentioned, 
the attributes 
of the SOC 
systems are avalanches 
reflected 
by the power-law pdf distributions. 
We follow 
with the construction 
of certain 
temporal characteristics 
by supposing their uncertain links to avalanches. 
By using Eqs.(\ref{signFtC1})-(\ref{signFtC3}) 
the evolution of any quantity 
can be mapped 
to the sequence of  passage times.
The example of this 
view represents pdf of $\tau_{F^C}$ 
depicted in Fig.\ref{Fig2}.
Because 
of the substantial 
difference 
in the exponents 
of pdf's belonging 
to $\tau_{F^C}$ and $\tau_{F^U}$, 
no universality attributes are
indicated.    
More encouraging  
should be to find of pdf's 
independent of the feedback type.
The natural way toward 
this aim  
seems 
to be the 
investigation 
of the passage time sequences  
linked to the order parameter 
of the canonical equilibrium 
of given system. 
In the case of 
the Ising model 
the ordering 
is described 
by the magnetization, 
or, eventually 
by isolated spin value.
Therefore, 
it seems to be logical 
to define the passage times 
$\tau_{m}$, 
$\tau_{s}$ 
given by Eqs.(\ref{signFtC1})-(\ref{signFtC3}) 
(corresponding to $A=m_L, s_i$, 
where $i$ 
is the arbitrary but 
fixed site position). 
The simulation 
results are depicted 
in Fig.\ref{Fig6}.
%--------- FIGURE ---------
%----------------------------------
In their structure, 
the following attributes 
relevant for interpretation 
in terms of SOC 
can be identified:
\begin{itemize}
\item[(I)] the power-law behavior
\begin{equation}
{\rm pdf}(\tau_{s}) \simeq \tau_{s}^{-\phi}\,,
\qquad 
\phi
\simeq 1.3
\label{expphi1}
\end{equation}
     with the 
     unique 
     exponent 
     $\phi$
     pertaining  
     to 
     different
     feedbacks
     $F_t^{C}$,
     $F_t^{U}$,
     i.e.
     to single-
     lattice 
     and
     two-lattice
     systems;
\item[(II)] 
     the
     interval 
     of dependence
     from~(I) 
     broaden 
     with 
     the size 
     of lattices
\item[(III)]the 
      exponent 
      $\phi$ 
      (at 
      the 
      present 
      level 
      of accuracy) 
      indistinguishable  
      for 
      pdf's taken 
      for sequences  
      $\{\tau_{s}^{ (k)} \}_{k=1}$ 
      and 
      $\{ \tau_{m}^{(k)}\}_{k=1}$. 
\end{itemize}
The high-temperature 
limit of pdf of 
$\{\tau_{s}^{(k)}\}_{k=1}$
can be easily 
derived
due to assumption about
the absence 
of spin-spin
correlations. 
Its form 
\begin{equation}
{\cal P}_{\tau_{s}}(p)
=
(1 - p)^{L^2\tau_{s}-1} p
\,,
\qquad
p=1/L^2\,
\label{Ptpara1}
\end{equation}
expresses 
invariance of
$s_i$
during
$(\tau_{s}-1/L^2)$
spin flips, and its
immediate change
after the 
$\tau$th spin flip 
occurring with 
probability 
$p$ of the random picking 
of $i$th site. 
The simulations 
carried 
for paramagnet
$T\gg T_{\rm c}$
depicted in 
Fig.\ref{Fig7}
agree 
with 
formula 
Eq.(\ref{Ptpara1}).

%--------- FIGURE ---------
Below $T_{\rm c}$ pdf 
splits into 
separable contributions 
fitted 
here by the bimodal 
distribution
$P_{\tau_{s}}=
 b_0 {\cal P}_{\tau_{s}}(p_0) +
 b_1 {\cal P}_{\tau_{s}}(p_1)$
with 
parameters
$b_0=0.003$,
$b_1=0.004$,
$p_0=0.61$,
$p_1=0.02$.
The supplementary
analysis
of statistics 
of the successive
time 
differences
of $m_L$
recovers 
that 
$b_1 {\cal P}_{\tau_{s}}(p_1)$ 
term 
originates
from the 
mechanism of the 
long-time 
tunneling among
nearly saturated 
states of the opposite polarity.
From the 
figure
it also follows 
that
power-law 
short-time 
regime described by 
Eq.(\ref{expphi1}) 
is formed 
only 
if the  
feedback 
mechanism is activated. 
This conditional occurrence of universality 
can be considered as an additional 
(IV)th attribute 
relevant for 
identification of SOC. 
The noncanonical equilibrium attained 
by the self-organized MC dynamics 
for $L=10$ 
leads to $P_{\tau_{s}}$ 
dependence, which can be approximated 
by the fit 
\begin{equation}
P_{\tau_{s}}
=
b_2\,
\tau_{s}^{-\phi}
\exp\left[
-
\left(
\frac{\tau_{s}}{\tau_2}
\right)^2
\right]
\end{equation}
with parameters 
$b_2=0.00264$,   $\tau_2=43.521$.

Let us to note that 
for 
the sand pile 
model \cite{Bak88} 
the spatial measure 
of {\em avalanche} 
is associated 
with the energy integral 
taken during 
the stage following 
{\em disturbance}. 
In the case of MC 
the analog 
of spatial measure 
can be the 
extremal magnetization  
\begin{equation}
m_{\rm max}^{(k)}
=
\begin{array}{c} 
\mbox{max}  
\\ 
 \mbox{
\tiny
 $t= t^{(k)}_{m}+1, 
 \ldots, 
 t^{(k+1)}_{m}$
 }
\end{array}
{\Big |} 
 m_{L_s,T_t} 
{\Big |}\,,
\label{Magneextre1}
\end{equation}
where $t_{m}^{(k)}= 
\sum_{j=1}^k 
\tau_{m}^{(j)}$.
The simulations show 
that pdf corresponding  
to sequence
$\{ m_{\rm max}^{(k)} \}_{k=1}$ 
can be approximated 
by ${\rm pdf}(m_{\rm max})
\propto 
(m_{\rm max})^{-\phi_{m}}$
with   
$\phi_{m}=-2.0\pm 0.1$ 
obtained for 
the narrow 
span $m_{\rm max}\in <0.2,\,0.4>$.
In agreement with 
SOC attribute 
labeled~I,
the independence 
of feedback type  
is identified.

\begin{figure}[!h]
\epsfig{file=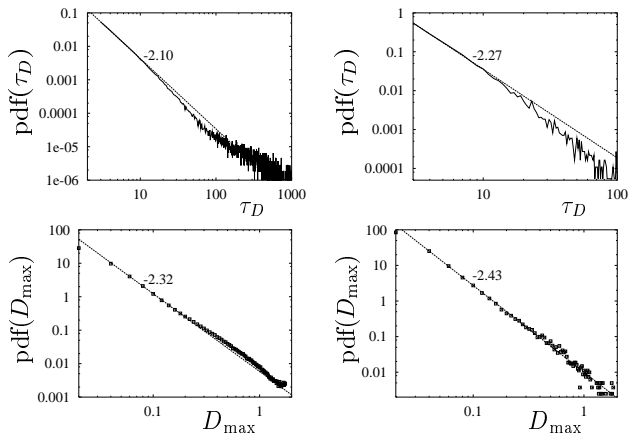,width=8.5cm,angle=0.0}
\caption{
The stationary 
power-law pdf 
distributions 
of the passage times  
$\tau_D$ 
calculated 
for different 
feedbacks (SOC criterion no.I)
$F^C_t$, $F^U_t$, 
self-organization 
parameters 
$\eta=10^{-4}$, 
$\Delta=10^{-5}$ 
and    
disturbance  
parameter 
$\epsilon_T=10^{-3}$
for sizes 
$L_1=10$, $L_2=20$.
}
\label{Fig8}
\end{figure}

It should be noticed 
that there remains 
the principal 
problem 
of the link 
between SOC
and presented 
self-organized 
MC method 
focused 
to the critical region. 
The problem is to 
identify
the specific 
algorithmic segment 
which should 
be interpreted
as an analogue 
of a disturbance 
initializing the avalanche. 
Fortunately, the advanced MC 
approach exists through which 
a disturbance 
can be absorbed  
into MC algorithm 
with minor 
violation 
of 
the original 
dynamics. 
In general, 
the approach 
of interest 
based on the 
coevolution 
of given system 
and its replica
is known under 
the term 
{\em damage spreading technique} 
\cite{Zheng98}.
To apply it, let us consider 
the self-organized 
MC referential system labeled 
here as $\{1\}_t$, 
which  
incorporates 
instant 
cumulants 
(single with $F^C_t$ 
or multi-lattice based on $F^U_t$), 
spin configurations 
and instant 
temperature 
$T_t^{\{1\}}$. 
Consider also
the replica 
counterpart 
$\{2\}_t$ 
of the system $\{1\}_t$.  
As is known, 
the parallel 
simulation of $\{1\}_t$  and 
$\{2\}_t$ should 
be applied with the identical 
pseudorandom sequences. 
Canonically, the measure 
of damage effect 
is then defined through the single-time 
two-replica difference
\begin{equation}
D^{\{1,2\}}_t=
m^{\{1\}}_{L_l,T_t^{\{1\}}} - 
m^{\{2\}}_{L_l,T_t^{\{2\}}}\,,
\label{distance1}  
\end{equation}
where 
$m^{\{1\}}_{L_l,T_t^{\{1\}}}$ 
and
$m^{\{2\}}_{L_l,T_t^{\{2\}}}$ 
are magnetizations 
of two lattices 
of the same size 
$L_l$ 
belonging 
to 
systems 
$\{1\}_t$ 
and 
$\{2\}_t$.
Using
Eqs.(\ref{signFtC1})-(\ref{signFtC3}) 
the sequence of differences 
$A_t\equiv D^{\{1,2\}}_t$, $t=1,2,\ldots$ 
is mapped 
onto the sequence 
of passage times 
$\tau_D^{(k)}$, 
$k=1,2,\ldots $.
In the 
case when 
$t$ coincides 
with one time 
between 
$ t_D^{(k)}= 
\sum_{j=1}^k \tau_D^{(j)} $,
the 
replica $\{2\}_{t_D^{(k)}}$  
is rebuilded 
within two steps:
\begin{enumerate}
\item
    all of the instant 
    cumulants 
    and spin 
    configurations  
    involved in 
    ${ \{ 2 \}}_t$ 
    are 
    replaced by 
    ${ \{1\} }_t$, i.e. $D^{\{1,2\}}_{t_D^{(k)}}=0$. 
\item
    the  
    replica 
    temperature 
    is 
    modified 
    according  
\begin{equation}
T_{t=t_D^{(k)}}^{\{2\}} 
= 
T_{t=t_D^{(k)}}^{\{1\}}
+
\epsilon_{T}\,, 
\end{equation}
where constant 
$\epsilon_T$ 
causes the small 
disturbance of 
the coincidence 
of the contents of  
$\{1\}_{t_D^{(k)}}$ and 
$\{2\}_{t_D^{(k)}}$.
\end{enumerate}
Evidently, 
the temperature 
disturbance 
plays role 
similar to 
adding of the grain to the sand pile. 
The sand pile 
is stabilized if the rest state 
occurring for
$t=t_D^{(k)}+\tau_D^{(k)}$  
is reached. 
The main idea of 
replica approach 
is that the relative motion 
of $\{2\}_t$ with respect to $\{1\}_t$ 
enhances 
the nonlinearity 
responsible for 
a wide range of 
$D_t^{\{1,2\}}$ 
responses to 
$\epsilon_T$. 
Thus, the only 
stochastic elements 
of replica 
simulation 
originate from the instants 
over which 
the 
content of
$\{2\}_{t=t_D^{(k)}}$ 
is replaced 
by the 
content of 
$\{1\}_{t=t_D^{(k)}}$.
In analogy 
to Eq.(\ref{Magneextre1}),
the complementary measure 
reflecting the spatial  
activity can be 
\begin{equation}
D_{\rm max}^{(k)}
=
\begin{array}{c} 
\mbox{max}  
\\ 
 \mbox{\tiny
 $t= t^{(k)}_{\rm D}+1, 
 \ldots, 
 t^{(k+1)}_D$
 }
\end{array}
{\Big |}  
D_t^{  \{ 1, 2 \}  }  
{\Big |}\,.
\end{equation}
Using this, the simulated path is  
mapped onto the sequence 
$\{ D_{\rm max}^{(k)}\}_{k=1}$.  
Consequently, 
the pdf's can be extracted which are depicted in Fig.\ref{Fig8}. 
They show 
that the effective exponents 
centred nearly $-2.3$ 
fit simulated pdf's fairly well. 
As in the case labeled~I,
pdf's related to  
$\tau_D^{(k)}$ and  
$D_{\rm max}^{(k)}$ 
are weakly susceptible 
to the feedback choice.   
Since both temporal 
and spatial 
power-law attributes 
of universality are indicated, 
the 
standard SOC paradigm 
can be considered 
as a framework 
adaptable for the analysis 
of the suggested 
self-organized 
MC dynamics. 

\section{Conclusions}

Several versions 
of MC algorithm
combining 
the self-organization
principles with 
the MC simulations 
have been 
designed.
The 
substantial 
feature of method
is the establishing of running 
averages coinciding with gamma 
filtering 
of the noisy 
MC signal.
The simulations 
are combined with 
the mean-field analysis
describing the motion
of temperature 
near to magnetic
transition point. 
The replica-based 
simulations indicate  
that pdf distributions 
of passage times
in a noncanonical 
equilibrium attain
the interval of the 
power-law behavior 
typical for the standard SOC 
pdf distributions.
We hope that the present  
contribution will 
stimulate further 
self-organized 
studies of 
diverse lattice models, 
e.g. those related to
percolation problem. 

\vspace{6mm}

{\bf Acknowledgement}

\vspace{4mm}

The authors would like 
to express their thanks to 
Slovak Grant agency 
VEGA~(grant no.1/9034/02) 
and internal grant VVGS 2003,  
Dept.of Physics, \v{S}af\'arik University, Ko\v{s}ice 
for financial support.

\vspace{5mm}

%--------------------------
 
   \end{document}